# Squeezed dual-comb spectroscopy

**Authors:** Daniel I. Herman,[1,*] Mathieu Walsh,[2] Molly Kate Kreider,[1,3] Noah Lordi,[3] Eugene J. Tsao,[1] Alexander J. Lind,[1] Matthew Heyrich,[1,3] Joshua Combes,[1] Jérôme Genest,[2,†] and Scott A. Diddams[1,3,‡]

**Affiliations:**

[1]Department of Electrical, Computer and Energy Engineering, University of Colorado Boulder; Boulder, Colorado, 80309, United States of America.

[2]Centre d'Optique, Photonique et Laser, Université Laval; Québec, Québec, G1V 0A6, Canada.

[3]Department of Physics, University of Colorado Boulder; Boulder, Colorado 80309, United States of America.

*Corresponding author. Email: daniel.i.herman@colorado.edu

†Email: jerome.genest@copl.ulaval.ca

‡Email: scott.diddams@colorado.edu

**Abstract:** Laser spectroscopy and interferometry have provided an unparalleled view into the fundamental nature of matter and the universe through ultra-precise measurements of atomic transition frequencies and gravitational waves. Optical frequency combs have expanded metrology capabilities by phase-coherently bridging radio frequency and optical domains to enable traceable high-resolution spectroscopy across bandwidths greater than hundreds of terahertz. However, quantum mechanics limits the measurement precision achievable with laser frequency combs and traditional laser sources, ultimately impacting fundamental interferometry and spectroscopy. Squeezing the distribution of quantum noise to enhance measurement precision of either the amplitude or phase quadrature of an optical field leads to significant measurement improvements with continuous wave lasers. But experiments demonstrating true metrological advantage of broad bandwidth squeezing with optical frequency combs are much less developed. In this work, we generate bright amplitude-squeezed frequency comb light and apply it to molecular spectroscopy using interferometry that leverages the high-speed and broad spectral coverage of the dual-comb technique. Using the Kerr effect in nonlinear optical fiber, the amplitude quadrature of a frequency comb centered at 1560 nm is squeezed by >3 dB over a 2.5 THz of bandwidth that includes 2500 comb teeth spaced by 1 GHz. Interferometry with a second coherent state frequency comb yields mode-resolved spectroscopy of hydrogen sulfide gas with a signal-to-noise ratio (SNR) nearly 3 dB beyond the shot noise limit, taking full metrological advantage of the amplitude squeezing when the electrical noise floor is considered. The quantum noise reduction leads to a two-fold quantum speedup in the determination of gas concentration, with impact for fast, broadband, and high SNR ratio measurements of multiple species in dynamic chemical and biological environments. Overall, this work establishes an understanding of quantum noise reduction in Fourier transform spectroscopy that will be foundational for further exploration with squeezed states in the broad application space of frequency comb metrology.

**Introduction**

Measurements of basic physical quantities including time, frequency and distance have been reshaped by a quarter century of optical frequency comb developments (*1–4*). Traceable, phase-coherent and broadband frequency combs have also found widespread use in atomic and molecular spectroscopy. For wavelength calibration of astronomical spectrographs, frequency combs push the precision of radial velocity measurements to levels required for the discovery of Earth-like exoplanets (*5*). As an active light source from the THz to the UV, combs extend the technological frontier for atomic and molecular science in both fundamental and applied spectroscopy (*6–8*). The development of high-repetition rate combs has allowed for fast biological and chemical sensing (*9, 10*). Low repetition rate combs (*11*) simplify the generation of coherent UV/X-ray light for high-resolution studies of high-energy quantum systems (*12*). With the rising diversity and relevance of comb-based applications, there is an increasing need to understand and extend the fundamental noise limitations on frequency comb measurements.

Recently, frequency comb metrology has reached the limit imposed by quantum mechanics in regimes of both bright coherent states and coherent states with small photon number (*13–15*). For the case of frequency comb spectroscopy, the standard quantum limit is set by the amplitude shot noise limit (SNL) of coherent state frequency combs (see Figure 1). To reach beyond standard quantum limits in frequency comb spectroscopy, one can realize comb architectures that incorporate squeezing as has been done previously for continuous wave (CW) lasers (*16–18*). Along these lines, several methods have been proposed for using squeezed frequency combs as spectroscopic tools (*19, 20*). These theoretical results imply that squeezing may yield a significant quantum advantage for frequency comb spectroscopy with realistic useful power levels (μW to mW) and attainable levels of squeezing (~10 dB). There has been significant progress in the generation of quantum combs using synchronously-pumped optical parametric oscillators, micro-resonator combs and other platforms (*21–24*). Quantum combs can be used as a multi-mode quantum resource for measurement-based quantum computation and quantum information processing (*25–28*). Also, quantum timing measurements using a sub-threshold synchronously-pumped optical parametric oscillator have demonstrated a small quantum advantage for microwatt level combs (*29*). However, a metrological advantage has yet to be demonstrated experimentally for squeezed comb spectroscopy.

Here, we describe a simple scheme for the generation and application of a bright amplitude-squeezed 1 GHz frequency comb for mode-resolved dual-comb spectroscopy (DCS) of molecular samples. Unlike existing proposals that utilize $\chi^{(2)}$ nonlinearities to generate squeezing over a few comb modes in a narrow band (*19, 20*), our method uses the $\chi^{(3)}$ nonlinearity to simultaneously suppress quantum noise for >2500 comb modes over a ~2.5 THz wide optical bandwidth. Our setup simplifies and improves upon well-established soliton squeezing techniques (*30–33*). Recent results using polarization soliton squeezing in fiber provided the first evidence of metrological advantage for interferometric phase measurements using the $\chi^{(3)}$ nonlinearity (*34*). We show that bright amplitude soliton squeezing provides a metrological advantage for broadband collinear dual-comb spectroscopy even when only one of the two frequency combs is squeezed. Overlapping the combs prior to the molecular absorption provides a baseline-calibrated transmission measurement that takes advantage of quantum correlations generated before the combining optics. Our measurement technique paves the way for future demonstrations of squeezed comb spectroscopy using nanophotonic platforms and opens the door for real-world uses for quantum-enhanced dual-comb spectroscopy.

**Quantum Noise in Dual-Comb Spectroscopy**

The DCS technique enables traceable spectroscopy across many terahertz of bandwidth at the resolution of the comb modes (teeth), but without moving parts involving variable delay lines or frequency-resolving dispersive optical elements (*7*). DCS works by interfering two combs with slightly different comb spacing to down-convert optical comb modes and optical absorption information to the radio frequency (RF) domain where the signal can be easily recorded and calibrated against an absolute frequency scale (see Figure 1b). Dual-comb interference appears as a multi-heterodyne RF comb in the frequency domain and as a periodic "interferogram" waveform in the time domain. The molecular absorption information is encoded as dips in the RF comb spectrum and as free induction decay (FID) signals in the time domain. DCS pushes the state-of-the-art for broad bandwidth, high-resolution, and traceable sensing across the electromagnetic spectrum with applications including open-path environmental sensing (*35*), chemical reaction monitoring (*36*), chemical imaging (*37*) and generation of accurate archival spectra for widely-distributed databases (*38*). Noise from the frequency comb sources and photodetectors places limits on the performance of dual frequency comb spectrometers by reducing the precision of the measured amplitudes of the RF comb teeth (see Figure 1b). For dual-comb spectroscopy, this noise is categorized into several sources including comb relative intensity noise (RIN), detector noise equivalent power (NEP), thermal noise and comb shot noise (*39*). At optical powers in the mW range, NEP and thermal noise can be negligible compared to other noise sources. For modern solid-state lasers with RIN at the -160 dBc/Hz level at MHz offset frequencies, the mW optical power regime is dominated by the fundamental quantum (shot) noise (*40*, *41*). Typically, the SNL is defined assuming shot noise is time stationary, which gives the shot noise current fluctuations in a 1 Hz bandwidth as $\sigma_{SNL} = \sqrt{2ei_{avg}}$ where $e$ is elementary charge and $i_{avg}$ is average photocurrent (*42*, *43*). We refer to this limit as the time stationary or coherent state SNL. It is possible to exceed this SNL in comb measurements (*e.g.* microwave generation, CW-comb heterodyne comparison) by taking advantage of the pulsed properties (cyclo-stationarity) of the frequency comb noise (*44*, *45*). Normally, dual-comb interferograms are low-pass filtered below half the repetition rate ($f_{rep}$) to include information only from the first "Nyquist window" (*7*). The SNL in this case is equivalent to the time stationary limit. Recent work has shown that it is possible to exceed the time stationary SNL in dual-comb interferometry by including information from the higher frequency "copies" of the interferogram above $f_{rep}/2$, which takes advantage of the cyclo-stationarity of the comb noise (*13*). Our work shows that quantum states of light can be utilized to exceed the time stationary SNL in dual-comb spectroscopy. Similar techniques should also surpass the cyclo-stationary limits in future demonstrations.

      Our method for quantum-enhanced dual-comb interferometry addresses properties of dual-comb measurements that previous theoretical approaches do not fully capture. In the time domain, the dual-comb interferogram can be decomposed into two regions: zero optical path difference (centerburst) and large path difference. In the centerburst region, the optical pulses from each comb are overlapped in time and therefore the measurement is mode-matched (see Figure 2a/b). The centerburst contains information about the spectral envelope of the two combs. Absorption by a molecular or atomic gas is observed as a FID signal that is mostly separated in time from the centerburst (*46*). From the quantum optics perspective (see Figure 2c), a classical dual-comb centerburst is generated by a beamsplitter interaction between a coherent state from the first comb and a mode-matched coherent state from the second comb, which results in a "quadrature-like" measurement (*47*). In this case, the phase and amplitude noise quadratures of

the comb fields are accessed via homodyne or heterodyne comparisons (*48, 49*), which will only provide a pure quadrature measurement in the limit that one comb is much stronger than the other (*50*). In the large optical path difference region, the two pulse trains are well-separated in time. This region can be modeled by coherent states from either comb interacting on a beamsplitter with a vacuum state which results in an "intensity-like" (photon number) measurement (*47*). Thus, in the large path difference regime the noise is purely derived from the quantum (shot) noise of the independent combs for any comb power ratio. Because the duty cycle (pulse length divided by pulse repetition period) of the 1 GHz comb used in the experiment is about 0.025%, the quantum noise of dual-comb interferometry is mostly dominated by intensity-like measurement noise.

In the intermediate FID region, the noise remains dominated by shot noise from an intensity-like measurement provided that the absorption depth and the corresponding FID ringing is small. This interpretation is consistent with established analyses of traditional Fourier transform spectroscopy (FTS). When compared to tunable laser spectroscopy, it is commonly acknowledged that FTS suffers a noise penalty due to shot noise from photons that do not interact with the absorbing medium (*51, 52*). We gain a metrological advantage in dual-comb spectroscopy by using a bright amplitude-squeezed comb to reduce the shot noise in the large path difference region assuming small absorption depths. Our interpretation is consistent with previous analyses that show a degraded quantum advantage in dual-comb with increasing loss (*19*). A future full quantum treatment of the centerburst region will enable improvements and extensions of quantum-enhanced dual-comb interferometry for additional impactful applications (*e.g.* ranging, time transfer, quantum state tomography).

**Experimental Setup**

A simplified schematic of the squeezed dual-comb spectrometer is shown in Figure 3a and more complete experimental details are included in the supplementary materials (SM). Here, we describe the generation and characterization of a single amplitude squeezed comb and the incorporation of this squeezed comb into a DCS setup. The optical frequency combs are generated using two commercial diode-pumped solid state femtosecond lasers (Menhir-1550) with 1 GHz repetition rates (*53*) emitting soliton pulses centered at 1563 nm (see Figure 3b). Here, direct passive mode-locking avoids active pulse regeneration and associated supermodes (*32*). The comb output is compressed close to the Fourier limit and sent to an un-balanced nonlinear Mach-Zehnder interferometer. This interferometer performs two important actions on the input soliton pulses: Kerr state generation and displacement of the soliton field. The pulse is split at the input of the interferometer into a strong pulse and a weak auxiliary pulse at 10:1 power ratio. The strong pulse is sent into a fiber with a large $\chi^{(3)}$ nonlinearity, ensuring that the strong pulse experiences a significant Kerr self-phase modulation. The Kerr effect applies an intensity-dependent phase shift to the pulse, which can be viewed in phase space as taking a symmetric coherent state to a Kerr state with a crescent-like distribution in the Wigner plane (see Figure 3c). Although this Kerr state is squeezed, the Kerr effect is photon-number conserving and therefore the squeezing angle is aligned such that an intensity noise measurement of an un-displaced Kerr state yields the same value as the input coherent state (*i.e.* the coherent state SNL) (*33, 54*). By slightly displacing the Kerr state in phase space the squeezing angle is shifted such that quantum noise reduction (QNR) or quantum noise amplification is observable in an intensity measurement. This small displacement is implemented with the Mach-Zehnder interferometer by recombining the strong squeezed state pulse and the weak auxiliary coherent state pulse (with an adjustable phase shift applied) on a variable beamsplitter at a 100 to 1 power ratio. If the relative phase between strong and weak pulses is approximately 90° or 270° the displaced state has nearly the same mean field as the original Kerr state but has observable amplitude squeezing or anti-squeezing, respectively (see Figure 3d). In the experiment, the Mach-Zehnder interferometer is implemented using the two axes of a polarization-maintaining highly nonlinear fiber (PM-HNLF) which improves common-mode phase noise rejection (*27*).

We designed a low-loss detection chain, including a high quantum efficiency (QE) photodetector with 300 MHz bandwidth, for single squeezed comb characterization and dual-comb spectroscopy measurements. Details about this detection chain are available in the SM along with additional results from an alternative detection chain based on a low QE photodetector with 5 GHz bandwidth. The high QE detection chain has an overall efficiency of ~78% including detector QE and optical loss. The detector photocurrent is converted to a voltage using a 50 Ohm passive transimpedance element and amplified using a low-noise voltage amplifier. For the single-comb squeezing analysis, this voltage output is characterized using an RF spectrum analyzer. A single photodetector is sufficient for direct detection of squeezing with the shot-noise-limited comb output being calibrated by RF spectrum analyzer measurements (see SM for details). The low noise floor of the analyzer allows the measurement of >3 dB of amplitude squeezing over a broad frequency range as described in detail in the next section.

In the dual-comb setup, the squeezed frequency comb is combined with a coherent state frequency comb using polarization optics to create a 50:50 beamsplitter (*55*). Before this combination a fraction of each comb is used for phase-locking to an optical phase reference that establishes mutual comb coherence. The locks are performed with digital electronics that can arbitrarily control the location of the dual-comb interferogram between 0 Hz and 100 MHz (*56*), and the repetition rate difference of the combs is set to 5.6 kHz. One output of the 50:50 beamsplitter is used as a spectral baseline reference measurement and the other output is sent through a gas cell containing pure hydrogen sulfide ($H_2S$) held at a pressure of 100 Torr. The baseline and signal interferograms, along with the optical phase reference tones are simultaneously recorded using a 1 GS/s, 16-bit digitizer card. The digitized interferograms are analyzed using a phase-correction algorithm (*57*, *58*) which is described in the dual-comb results section. In this demonstration, the squeezed comb is about 1000 times stronger than the coherent state comb due to technical limitations (*e.g.* ADC dynamic range and photodetector nonlinearity). In the future, these limitations can be overcome through further engineering of the detection chain or by adjusting the squeezing procedure to operate at slightly lower powers.

**Single Comb Squeezing**

Our setup has several differences over previous Kerr squeezing experiments (*32*, *33*) which collectively allow for simpler generation of amplitude squeezed combs. We implement the squeezing in PM-HNLF which has a high nonlinear coefficient similar to photonic crystal fiber (PCF) (*59*) or single-mode HNLF (*60*), but provides a high polarization extinction ratio and low-loss coupling appropriate for the ~20 pJ pulse energy of the 1 GHz comb. The high nonlinearity and low insertion loss of the PM-HNLF helps avoid external optical amplification despite the low pulse energy of the 1 GHz comb. The 1 GHz optical sampling of this comb source fully resolves noise associated with guided acoustic wave Brillouin scattering (GAWBS) in the PM-HNLF without aliasing (*32*, *50*, *61*). Thus, broadband squeezing in the RF domain is easily quantified between presumed GAWBS peaks at ~23 MHz and ~68 MHz (see Figure 3e). The maximum squeezing in our setup occurs when 17 mW of the strong pulse is out-coupled from the PM-HNLF. The most squeezing is measured when 14.7 mW of the strong squeezed pulse is recombined with ~150 μW of the weak auxiliary pulse. Without the auxiliary pulse, the measured photocurrent intensity noise matches that of a coherent state comb without the PM-HNLF (held at the same optical power). This observation confirms that the Kerr effect is photon number conserving and therefore does not allow for sub-shot noise measurements without displacement of the Kerr state.

With the auxiliary pulse and the high QE detector, we measure 3.8 dB of squeezing and 4.0 dB of anti-squeezing from 45 MHz to 50 MHz and >3 dB squeezing in a broader window from 10 MHz to 100 MHz. The maximum squeezing level is comparable to similar demonstrations of nonlinear Mach-Zehnder interferometers (*33, 59*). The low QE detector has greater electrical bandwidth and with it we measure squeezing from 20 MHz to 500 MHz at the ~2 dB level (see SM for further details). This result is among the broadest reported single comb squeezing, covering a complete Nyquist window (*i.e.* up to $f_{rep}/2$) (*62, 63*). Generally, conventional quantum sensing applications have utilized either broadband low frequency squeezing (*e.g.* LIGO) or narrow-band high frequency squeezing (*64, 65*). While there are possible uses of broadband high frequency squeezing for quantum information processing (*66*), our results demonstrate that DCS is a compelling metrology application for bright broadband squeezing. For example, with a 5 kHz repetition rate difference, the 500 MHz electrical bandwidth could support >90 THz of quantum-enhanced DCS.

**Collinear Quantum-Enhanced Dual-Comb Spectroscopy**

The time domain quantum optical picture of dual-comb interferometry makes it clear that bright squeezed states can be used to gain a metrological advantage in DCS. However, we must also consider which dual-comb readout schemes are compatible with the use of bright amplitude-squeezed states. A recently proposed approach for quantum-enhanced dual-comb employs two-mode squeezing between a comb tooth and its sidebands and places the absorbing medium in the path of one comb before combining on a beamsplitter, *i.e.* asymmetric dual-comb (*7, 19*). While possibly useful, this two-mode squeezing scheme is not trivial to achieve using off-the-shelf components and has not been experimentally demonstrated. In our scheme, one of the combs is in a bright state with single-mode amplitude squeezing. Bright single-mode amplitude squeezing of optical pulses is a well-established technique and can be performed using simple procedures (*33*). After mixing this squeezed comb with a coherent state comb using a beam splitter interaction, the output interferograms will have opposite signs and the quantum noise will be partly anti-correlated. The optimal measurement strategy for asymmetric dual-comb involves subtracting the photocurrents through balanced photodetection, but this operation would both maximize our signal and destroy any advantage from squeezing.

Instead, we place the absorbing medium in the path of both combs after they are combined (*i.e.* collinear dual-comb). In this case, the absorption signal is imprinted on one interferogram and the other interferogram is used as a spectral baseline reference (see Figure 3A). It is shown in the SM that the complex spectral division of two correlated shot noise processes (*i.e.* time domain photocurrents) retains the correlations of the individual processes and produces an enhanced estimate of molecular transmittance. Importantly, the time domain correlations generated by the squeezing operation remain intact in the spectral domain due to the linearity of the Fourier transform. Furthermore, in the limits of high SNR and low absorbance, spectral division treats anti-correlated noise similarly to the time domain addition of the photocurrents, which is the standard method for characterizing amplitude squeezed states (*67–69*). This addition operation can be performed electronically with either analog methods or digital post-processing. Here, we choose to digitize the two interferograms separately for easy comparison of time domain addition and spectral division. Adding the two photocurrents cancels the opposite sign interferograms and is in that sense similar to optical techniques used to reduce the centerburst dynamic range (background suppression) in FTS (*70*) or DCS (*71*). In our experiments, the noise is also anti-correlated and is thus also partially cancelled by the photocurrent addition.

We generate dual-comb interferograms between our amplitude squeezed comb (~15 mW) and a weak (~10 µW) coherent state comb. Even with unbalanced comb powers, we receive high quality DCS data with quality factor near the state-of-the-art ($>10^7$ Hz$^{-1}$) (*72*). We set the optical lock frequencies such that the interferogram is centered at 50 MHz and covers about 25 MHz of RF bandwidth. The centerbursts from the two channels are nearly identical except for a 180° phase shift (see Figure 4a). As an initial test, the raw interferograms are temporally aligned using a cross-correlation and then summed (see Figure 4b/c). The quantum noise at large path differences is analyzed for the no auxiliary pulse (SNL), squeezed (SQZ) and anti-squeezed (ANTI-SQZ) cases. Without accounting for the electrical noise floor of the high-speed digitizer, the QNR is 55% (2.6 dB) in the time domain. When we subtract the noise floor, the squeezing level is calculated at 3.5 dB. Similar QNR is obtained if the SNL is derived by either adding the photocurrents in the absence of the auxiliary pulse or by subtracting the squeezed photocurrents. These results confirm the single comb squeezing measurements and show that the squeezing is useable in a dual-comb context.

**Frequency Domain Dual-Comb Results**

Mode-resolved DCS with quantum enhancement requires phase correction of the interferograms, followed by Fourier transformation and spectral division (see Figure 4d). The phase correction removes both interferogram phase fluctuations and timing jitter using a process that is now a standard technique in dual-comb spectroscopy and therefore only the general structure of the correction is reviewed in this work (*57*). First, a fast phase correction is implemented on both interferogram channels using the digitized phase reference signals. This correction step removes phase noise at offset frequencies up to ~1 MHz. Next, a slow phase correction and resampling is performed on the interferograms based on the phase calculated from the interferograms themselves (*58*). After phase correction, the interferogram trains from both channels are Fourier transformed to generate mode-resolved spectra (see Figure 5). The squeezing is visible in the wings of the spectrum and between the individual RF comb lines for the individual channels although at a diminished level (~1 dB) since each channel sees 50% extra loss when viewed independently. The phase-corrected interferograms are averaged to generate one high SNR interferogram per channel and the two channels are phase-aligned using a cross-correlation. The full QNR is revealed after the interferograms are Fourier transformed and spectra are divided to yield a transmission spectrum. Remaining variations in the spectral baseline are corrected using a cepstral-domain filtering process, which simultaneously removes etalons and noise at extremely large path length differences (*73*). We fit the transmittance to a model constructed from the HITRAN2020 database (*74*). Our fit uses a standard nonlinear least squares regression to extract path length information with pressure, temperature and concentration as input parameters. We perform the fit over a 20 nm bandwidth centered at 1563 nm which covers 67 rovibrational transitions of $H_2S$ with spectral line intensity above $10^{-23}$ cm$^{-1}$/(molecule·cm$^{-2}$) (see Figure 6a/b). The low pressure of our cell (100 Torr) gives a linewidth of about 1 GHz and thus only 3 or 4 comb lines sample any given absorption feature (see Figure 6c). Nonetheless, our data is a near perfect match to the HITRAN2020 model with residuals in the sub percent regime in <10 ms of averaging, validating the performance of our phase-correction and baseline removal algorithms.

As seen in Figure 7, the fit residuals average as $\sqrt{N_i}$ in all three conditions (SNL, SQZ, and ANTI-SQZ) where $N_i$ is the number of averaged interferograms. A $\sqrt{N_i}$ power law is fit to each of the three conditions to reveal the reduction in noise as function of number of averages. The squeezing or anti-squeezing acts to change the constant multiple factors on this $\sqrt{N_i}$ power law (*16*). The ratio of the constant multiple factors between the SQZ and SNL cases is 0.51 representing a 2.9 dB improvement in power SNR (equivalently a 2× quantum speed up), nearly the same result seen in the raw time domain data. Although our data was highly processed to yield the baseline-corrected fits, none of the processing steps masked the QNR. Thus, fully-locked combs are not necessary for quantum-enhanced spectroscopy. We also note that the squeezing did not bias the fit result which also showed improved precision matching the QNR in the time and frequency domains.

**Outlook**

In this paper we have applied advances from quantum optical sensing to the powerful technique of Fourier transform spectroscopy with dual frequency combs. This innovation has allowed us to realize the first quantum-enhanced mode-resolved DCS over an unprecedented 2.5 THz optical bandwidth with 2.9 dB QNR below the SNL. Our simple and robust implementation employs bright Kerr squeezing and a collinear spectroscopic interaction to realize significant metrological advantage that was not predicted in recent theoretical proposals. Consequently, we anticipate this work will stimulate future research to test the limits of quantum advantage in broadband frequency comb metrology and spectroscopy. For example, it will be useful to evaluate other DCS geometries over a range of comb powers, and combine DCS with larger amounts of squeezing using different platforms. While it is well known that Raman effects in fibers limit the achievable squeezing (*75*, *76*), these effects may be less deleterious to squeezing in a nanophotonic platform (*e.g.* silicon nitride (*77*)).

Furthermore, our work provides the foundation from which we can investigate even greater increases in the SNR achievable with quantum-enhanced dual-comb spectroscopy. In order to maximize the SNR, the two combs should be power-matched. However, this poses a challenge for digital processing, in that a single ADC cannot record both the centerburst signal and quantum-limited electrical noise floor at the same time. One path to overcome this limitation would involve combining multiple ADCs to yield a wider dynamic range (*78*). With an improved ADC setup and power-balanced frequency combs, the largest possible SNR in a collinear DCS experiment results from squeezing both combs, as we outline in the SM. With two power-matched squeezed combs, a collinear experiment should be able to achieve QNR at the greatest reported level of Kerr fiber squeezing (~7 dB) (*75*).

Because of the tremendous breadth of applications of frequency combs, we expect that squeezed dual-comb spectroscopy will have a number of immediate use-cases including improved characterization of fast reaction kinetics (*36*), reduced power levels for spectro-imaging of biological samples (*37*, *79*) and improved SNR to overcome detector nonlinearity issues (*72*). While our present measurements extend over 2.5 THz, many spectroscopic applications will benefit from still broader optical bandwidth of squeezed states, as has been recently demonstrated with nanophotonic technology (*80*). As a variation of the Fourier transform spectroscopy (FTS) technique with improved speed, sensitivity and frequency axis traceability, dual-comb presents numerous opportunities for trace gas sensing, spectro-imaging and industrial analysis. Although FTS has begun to harness single photon states of light (*81*), no FTS technique has yet taken advantage of bright squeezed states of light. Here, our insights into the quantum noise of dual-comb spectroscopy will push the state-of-the-art for quantum FTS as a whole. Finally, we expect this work will inform future endeavors aimed at improved frequency comb heterodyne and homodyne techniques for impactful applications in distance metrology, optical clock comparison, and optical time transfer.

**Acknowledgments:** The authors acknowledge helpful technical discussions with Jean-Daniel Deschênes, Curtis Rau, Shawn Geller and Akira Kyle. The authors thank Jun Ye and Nathan Newbury for valuable comments on the manuscript. The authors also thank Kevin Cossel, Ian Coddington and Tsung-Han Wu for loaning technical equipment.

**Funding:** The authors acknowledge funding from the NSF QLCI Award No. OMA-2016244 and ONR N000142212438. E.J.T. acknowledges support from the NDSEG Fellowship and M.K.K and M.H. acknowledge support from the NSF Graduate Research Fellowship Program. M.W. and J.G. acknowledge support from NSERC and Photonique Quantique Québec (PQ2).

**Author contributions:** Conceptualization: DIH, MW, MKK, NL, EJT, JC, JG, SAD. Methodology: DIH, MW, MKK, AJL, MH, JG, SAD. Software: DIH, MW, MKK, JG. Formal analysis: DIH, MW, MKK, NL, JC, JG, SAD. Investigation: DIH, MKK, MW, JG, SAD. Resources: JG, SAD. Writing – original draft preparation: DIH, MKK, JG, SAD. Writing – review and editing: All authors. Visualization: DIH, MKK, MW, JG, SAD. Supervision: JC, JG, SAD. Project administration: SAD. Funding acquisition: SAD.

**Competing interests:** The authors declare no competing interests.

**Data and materials availability:** All data used to generate the figures and main results of this work are available upon reasonable request.

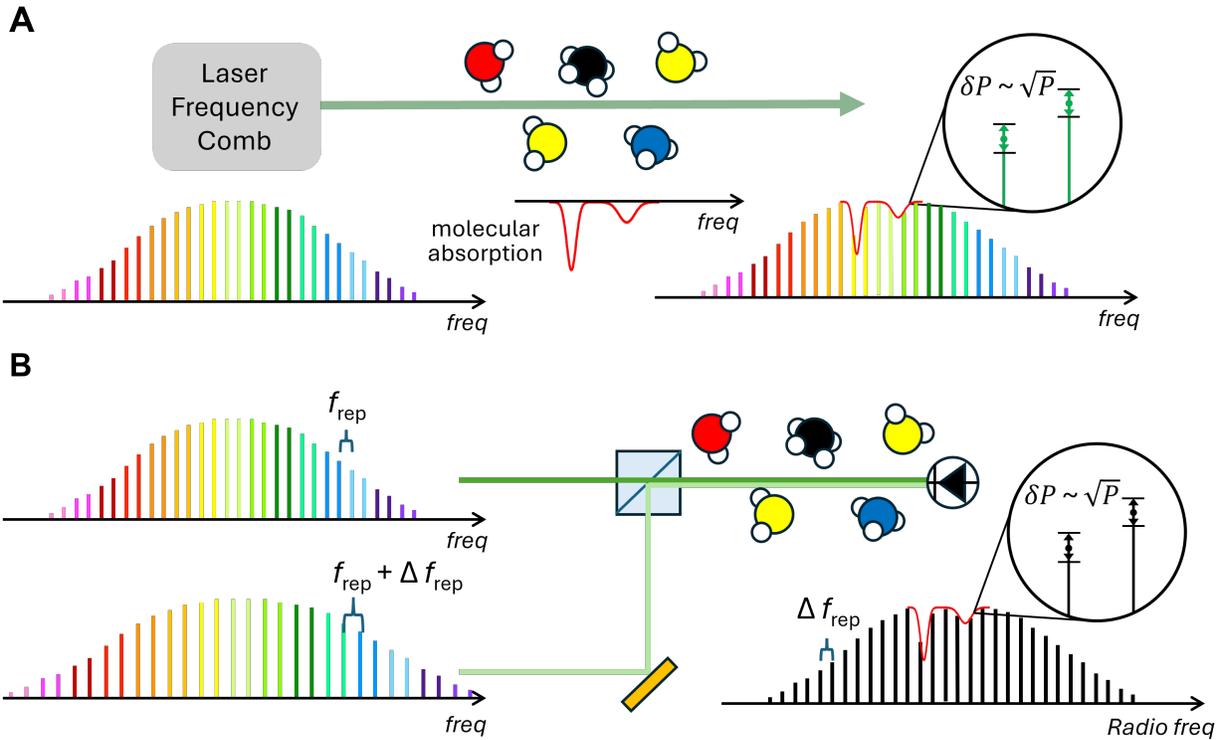

**Fig. 1. Frequency comb spectroscopy and quantum noise.** (A) Schematic of direct frequency comb spectroscopy. Molecular absorption is imprinted on the comb and the ultimate quantum sensitivity limit is set by amplitude shot noise ($\delta P \sim \sqrt{P}$) on the individual comb teeth. (B) The interference of two combs with a difference in repetition rates ($\Delta f_{rep}$) maps optical frequencies to radio frequencies in a one-to-one manner. The Fourier transform of the photodetector output generates a radio frequency comb signal. In this situation, the ultimate sensitivity limit is set by shot noise on the radio frequency comb teeth.

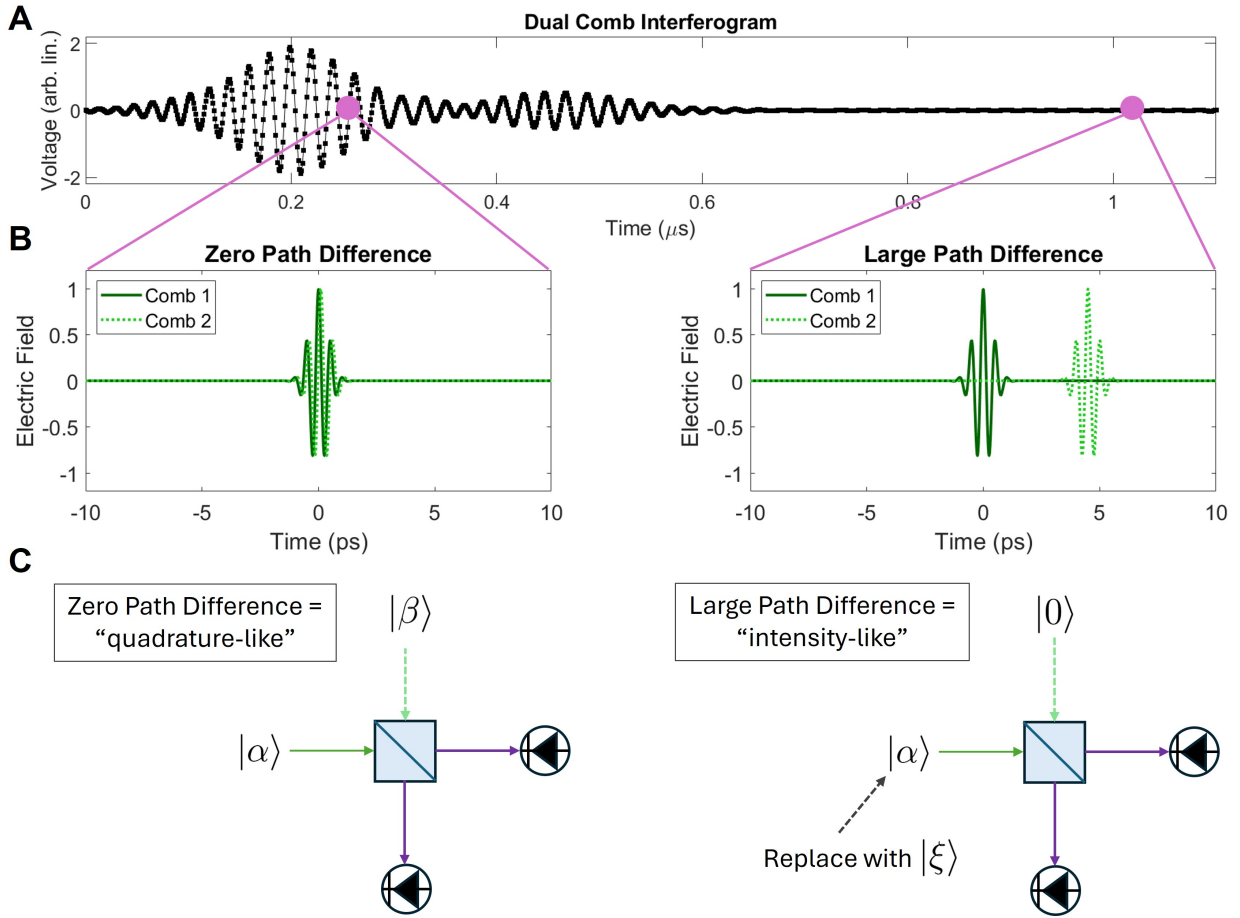

**Fig. 2. Time domain quantum measurement picture of dual-comb interferometry.** (A) An example of a digitized dual-comb interferogram in the time domain. X-axis is lab time in microseconds. Each point in the interferogram arises from the integrated interferometric overlap of two subsequent pulses from the dual-combs. This point is illustrated further in (B) where we show a simplified model of electric fields generating the dual-comb interferogram from zero path difference (left) and large path difference (right) cases. Here, the X-axis is effective optical path separation time in picoseconds. (C) Simplified single-mode picture of quantum measurement in two extreme cases. For zero path difference, the measurement obtains a quadrature-like character that one would receive from interfering two coherent states of magnitude α and β on a beamsplitter. In the large path difference case, the measurement has an intensity-like character which amounts to interfering a coherent state with a vacuum state (from the perspective of either comb). By replacing the coherent state with an amplitude squeezed state of magnitude ξ we can achieve quantum noise reduction in a dual-comb measurement.

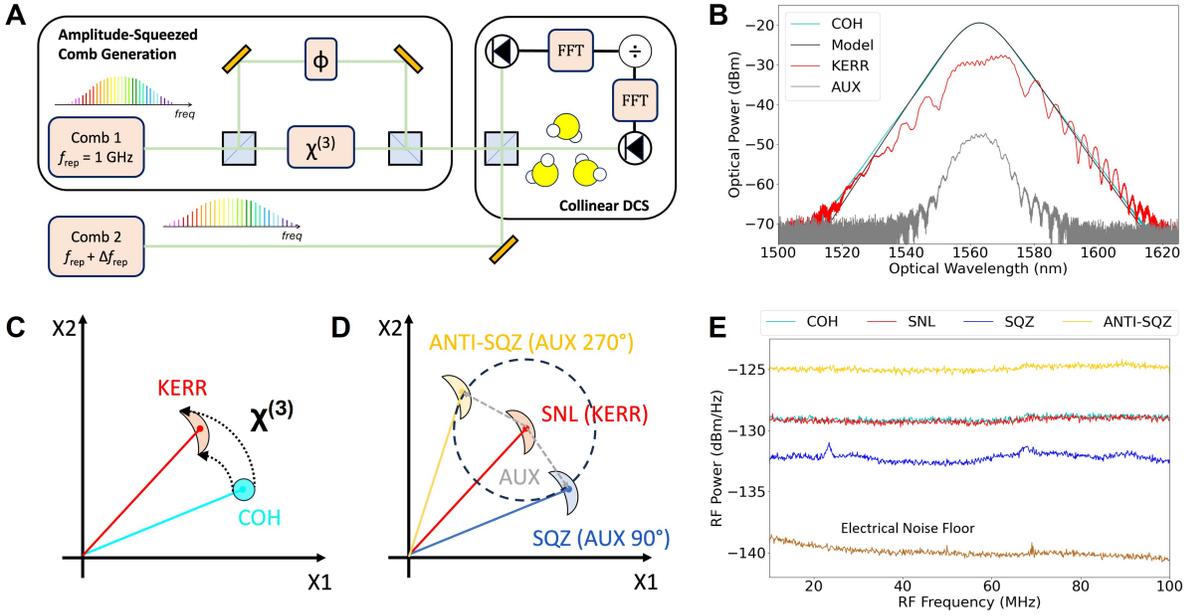

**Fig. 3. Dual-comb schematic and single-comb squeezing concept with results.** (A) Simplified schematic of quantum-enhanced dual-comb spectroscopy setup. A nonlinear Mach-Zehnder interferometer is used to generate an amplitude-squeezed frequency comb. The amplitude-squeezed comb is then combined with a coherent state frequency comb. Molecules are placed in one of the dual-comb beams. Each dual-comb signal is detected, digitized and then Fourier transformed (FFT). The outputs are then divided to generate a quantum-enhanced normalized transmission spectrum. (B) In light blue, the optical spectrum of the coherent state frequency comb (COH) before squeezing along with a soliton model spectrum (black). In red, optical spectrum of the strong squeezed pulse (KERR) after propagation through PM-HNLF. In gray, optical spectrum the weak auxiliary pulse (AUX) after propagation on the orthogonal axis of the PM-HNLF. (C) Phase space diagram for generation of squeezed states using the Kerr nonlinearity. An amplitude-dependent phase shift takes a coherent state (COH) and generates a tilted crescent shaped distribution (KERR) in phase space. X1 and X2 represent the bosonic quadrature operators, analogous to position and momentum. (D) Phase space diagram of displacement using the auxiliary pulse. By displacing the KERR state (SNL) with an auxiliary pulse of the right magnitude and phase, we can produce an amplitude anti-squeezed state (ANTI-SQZ) or an amplitude squeezed state (SQZ). (E) Electrical spectrum of photovoltage noise of coherent state (COH), Kerr state (KERR), auxiliary-displaced squeezed state (SQZ), auxiliary-displaced anti-squeezed state (ANTI-SQZ) and the electrical noise floor of the detection chain (brown). All traces are measured using a single photodetector with quantum efficiency of 95% and spectrum analyzer resolution bandwidth = video bandwidth = 1 kHz. See the SM for a schematic of the single-comb squeezing characterization setup.

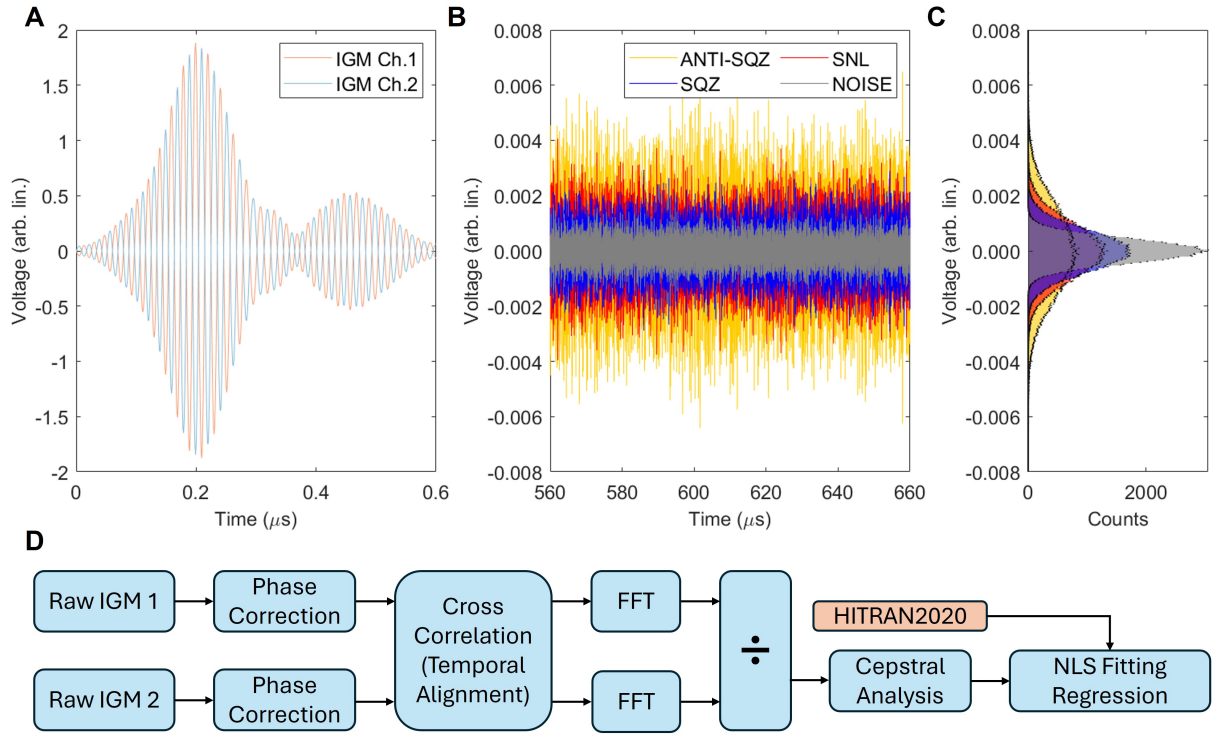

**Fig. 4. Dual-comb time domain data and mode-resolved spectroscopy processing chain.** (A) Nearly identical interferograms generated using the two photodetection channels on each side of the combining beamsplitter shown in Fig. 3A. Note the 180° phase shift between the two interferograms. The interferograms have been digitally filtered (center frequency 50 MHz; bandwidth 20 MHz). (B) Interferogram signal at large path difference after digital addition of two detector signals under four experimental conditions. Shot-noise limit (SNL) is the un-displaced Kerr state and the displaced SQZ and ANTI-SQZ states are shown along with the electrical detection chain noise floor in gray. A QNR of 2.6 dB is estimated between SNL and SQZ. (C) Histogram of voltage noise from (B) clearly showing QNR and quantum noise amplification of SQZ and ANTI-SQZ states. (D) Simplified block diagram detailing processing chain to generate quantum-enhanced mode-resolved dual-comb spectra and the subsequent fitting routine. Note: FFT = fast Fourier transform; NLS = nonlinear least squares; IGM = interferogram.

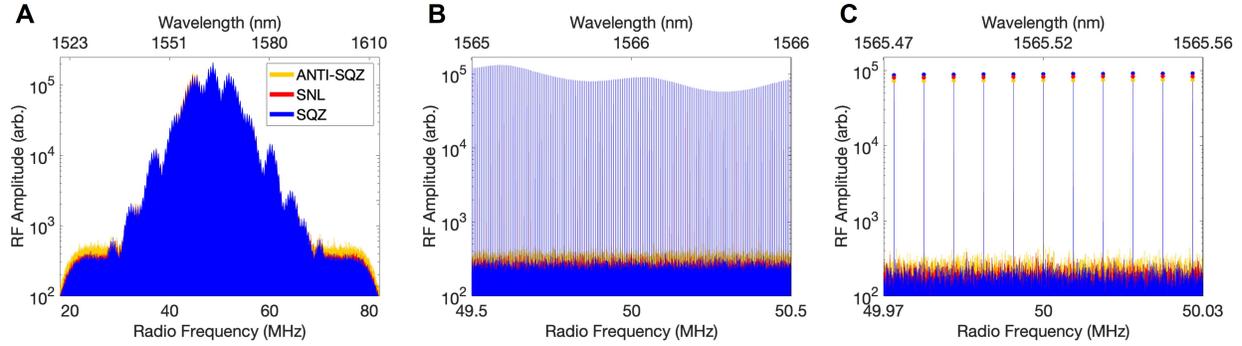

**Fig. 5. Mode-resolved squeezed dual-comb spectroscopy.** (A) Radio-frequency spectrum of phase-corrected dual-comb spectroscopic data from a single photodetection channel showing the dynamic range and full equivalent spectral bandwidth of 60 nm or 7.3 THz. These data are obtained from averaging 1112 interferograms. Squeezing of ~1.0 dB is visible between SQZ and SNL. (B) A zoomed in view to 1 MHz bandwidth clearly shows the mode-resolved nature of the dual-comb spectrum. The sinusoidal amplitude modulation is due to an etalon originating from a neutral density filter. (C) Additional zoom in to 60 kHz bandwidth showing QNR between the RF comb teeth. The markers indicate the peaks of the comb teeth to illustrate the improved signal-to-noise ratio with squeezing.

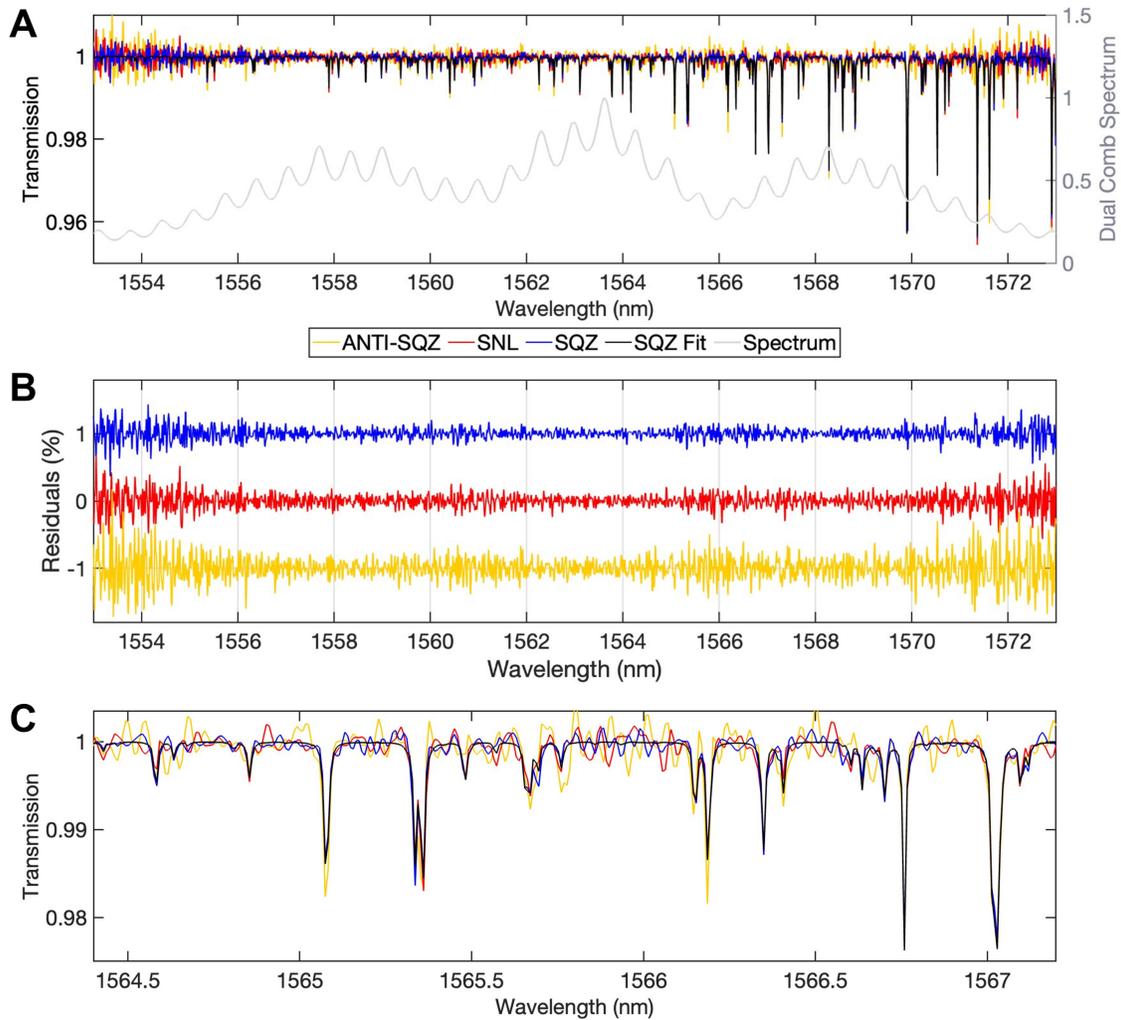

**Fig. 6. Dual-comb transmission spectrum and fits.** (A) Averaged dual-comb transmission spectrum generated from the ratio of the Fourier transform of each individual channel. These data are obtained from the averaging of 1112 interferograms or ~198 ms of data acquisition. The narrow $H_2S$ absorption features are most clearly visible in the black curve, which is a fit to the SQZ dataset. In black, a fit to the SQZ dataset is shown. In gray, a dual-comb spectrum from one of the channels is shown for reference. (B) Residuals (transmission minus fit) are shown for three cases: SNL (red), SQZ (blue) and ANTI-SQZ (yellow). For this averaging condition, the QNR was ~2.6 dB. The variation in the noise of the residuals results from the uneven power in the dual-comb spectrum that is shown above. (C) Zoom-in on several $H_2S$ absorption lines showing the high-quality nature of the data and fitting routine.

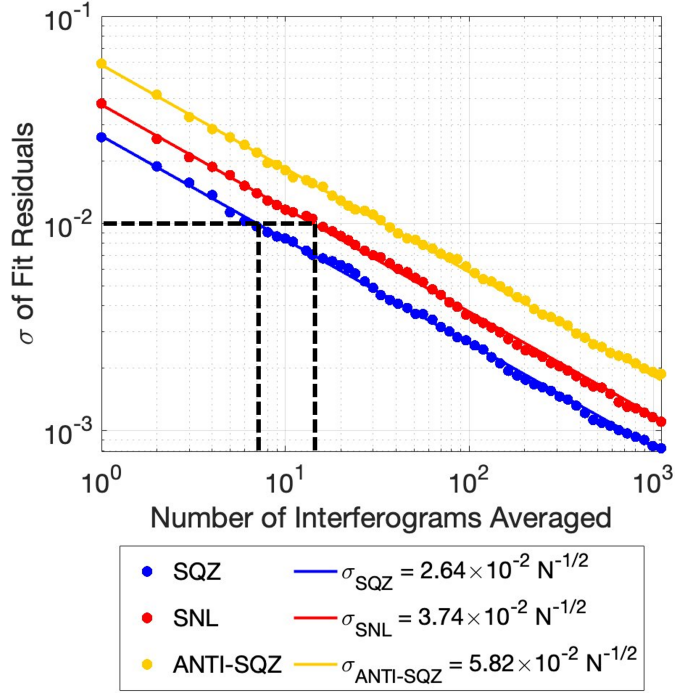

**Fig. 7. Quantum speed up with a squeezed frequency comb.** The standard deviation $\sigma$ of the residuals of the fit to spectroscopic absorption model is plotted versus the number of interferograms averaged ($N$). All three cases (SNL, SQZ and ANTI-SQZ) show the same inverse trend with $\sqrt{N}$, but with different offsets. The ratio of the calculated slopes for the SQZ and SNL cases is approximately 1.42 (or $\sim\sqrt{2}$). As shown by the dashed lines, this leads to a quantum speed-up of 2 ×, meaning that the same precision of $\sigma = 10^{-2}$ is achieved with one-half the number of averages when the squeezed comb is employed.

**Supplementary Materials**

Supplementary Text

Materials and Methods

The combs in this work are generated using two commercial diode-pumped solid state femtosecond lasers (Menhir-1550) with 1 GHz repetition rates (*53*). Both combs have polarization-maintaining (PM) fiber-coupled outputs and each emit ~50 mW of light in ~400 fs pulses. Their spectral envelopes are nearly ideal hyperbolic secant shapes characteristic of fundamental optical solitons. The soliton spectra are centered at 1563 nm with a 3 dB optical bandwidth of 12 nm. One of the combs is chosen to be the squeezed source. Before applying nonlinearity to this comb, its pulses are compressed using PM normal dispersion fiber (nLight LIEKKI Passive-4/125(-PM), 0.22NA) to 260 fs corresponding to a time-bandwidth product of 0.38. The PM normal dispersion fiber is coupled into free space and launched into a birefringence compensator (*33*, *82*) which splits the comb into two separate pulses at a power ratio of 10 to 1. The strong and weak pulses are then recombined with orthogonal polarizations. The phase of the weak pulse is adjustable using a mirror (Edmund Optics #64-022) mounted on a piezo-electric transducer (PZT) with a 3.6 μm throw (Thorlabs PA4FKW). The two pulses are coupled onto the orthogonal axes of PM highly nonlinear fiber (HNLF) with an extinction ratio of ~23 dB. The PM-HNLF (OFS HNLF-PM) has a dispersion of ~5.6 ps/nm/km, nonlinear coefficient of 10.7 $W^{-1}·km^{-1}$ and a length of 7 m. This length corresponds to approximately 1.5 soliton periods.

  The strong pulse experiences a large amount of Kerr nonlinearity resulting in a highly self-phase modulated spectrum at the PM-HNLF output. By balancing the Kerr nonlinearity with fiber dispersion, the strong pulse undergoes soliton squeezing resulting in a crescent-shaped distribution in phase space. It is well known that the Kerr nonlinearity is photon-number conserving and consequently there is no reduction in amplitude noise for an undisplaced Kerr state (*33*, *54*). Using a well-established method (*33*), the two axes of the PM-HNLF are utilized as a Mach-Zehnder interferometer which enables co-propagation of our strong squeezed pulse alongside the weak "auxiliary" pulse, reducing any un-wanted effects from non-common phase noise. The two pulses are recombined after the PM-HNLF at a 100 to 1 power ratio using a half-wave plate and a polarizing beam splitter. By shifting the phase of the auxiliary pulse, we displace the Kerr state to reveal either squeezing or anti-squeezing (see Figure 3c). the polarization of the displaced Kerr state is purified using an additional polarizing beamsplitter and then the squeezing is evaluated using direct detection by a singular photodiode (see Fig. S1).

  Our squeezed comb source is characterized with two different photodetectors in order to measure both the broadest possible squeezing and the largest possible amount of squeezing for our nonlinear optical setup. The first is a fiber-coupled, 5 GHz InGaAs photodiode with a quantum efficiency (QE) of 72% (Thorlabs BDX1BA). This detector has been previously used in conjunction with the 1 GHz solid-state frequency combs to generate shot-noise limited dual-comb spectra at high power levels (~7 mW) as well as linear dual-comb spectra with fiber laser combs up to 30 mW (*40*, *72*). The optical throughput between the PM-HNLF and the detector is 82% and the fiber-coupling efficiency is ~90% so the total detection efficiency in this case is 53%. The second detector is a free-space coupled 300 MHz InGaAs photodiode with an anti-reflection (AR) coating optimized for QE = 95% at 1550 nm (Fermionics Opto-Technology FD500N-1550). Both photodiodes are connected directly to a bias-tee (Mini-Circuits ZFBT-4R2GW+) and the RF output is sent through a 50 Ohm pass-through connector that acts as a passive transimpedance element. The high-speed detector is low-pass filtered at 520 MHz (Mini-Circuits BLP-550+) and the high QE detector is low-pass filtered at 98 MHz (Mini-Circuits SLP-100+). The voltage signal is amplified using a low-noise RF amplifier (RFBay LNA-725). The amplifier output is low-pass filtered again and sent to an RF spectrum analyzer (Siglent SSA 3021X). A shot noise calibration curve for the high quantum efficiency diode is shown in Figure S2. This curve was measured by applying known optical powers to the photodiode and calculating the shot noise level based on the manufacturer's responsivity number along with the measured RF gain in the detection circuit and a 50 Ohm load (*43*). The measured noise floor of the circuit was also added as a parameter in the fit model.

In the dual-comb setup, the squeezed frequency comb is combined with a coherent state frequency comb using two polarizing beamsplitters and a half waveplate to create a 50:50 beamsplitter (*55*). Before this beamsplitter, a small amount of each is picked off for locking to a narrow linewidth CW diode laser centered at 1564 nm (Redfern Integrated Optics PLANEX). The optical heterodyne beats between the combs and the CW laser are detected at high signal-to-noise ratio using balanced photodetectors (Thorlabs PDB130C-AC). The combs are locked to the CW laser using feedback to fast and slow PZTs with ~750 milliradians of integrated optical phase noise (from 3 kHz to 5 MHz). The locks are performed with digital phase lock boxes (Waxwing Instruments CLB) that allow arbitrary control of the dual-comb interferogram center frequency between 0 Hz and 100 MHz (*56*). The repetition rate difference of the combs is set to 5.6 kHz. One 50:50 beamsplitter output is used as a spectral baseline reference measurement the other output is sent through a gas cell containing pure hydrogen sulfide ($H_2S$) held at a pressure of 100 Torr (Wavelength References H2S-T(12x10)-100-1550)). The cell has AR-coated wedged windows, a total optical transmission of 95.9% and an optical path length of 10 cm. For the dual-comb measurements, we choose to use the higher efficiency photodetectors to reveal the maximum squeezing level. After the same detection chain used in the single comb experiments, we record the reference and signal interferograms using a 1 Gs/s, 16-bit digitizer card (GaGe CSE161G4-LR). The digitized waveforms are then analyzed using a phase-correction algorithm (*57*, *58*) described in the main text.

Discrete Spectrum of Correlated Gaussian Random Processes

For propagation distances restricted to a few soliton lengths, the quantum state generated by $\chi^{(3)}$ nonlinearities can be approximated by a quadrature squeezed state (*83*), where the non-gaussian properties of the quantum state are negligible (*84*). Quadrature squeezed states are represented by two dimensional gaussian distributions in the Wigner plane. A beamsplitter interaction between a quadrature squeezed state and a vacuum state outputs two correlated gaussian states (*85*). Assuming a large photon number for the squeezed state (bright squeezed state), the photo-currents generated by capturing each beam splitter output on a separate detector have statistics that are approximated by correlated gaussian distributions (*86*).

Considering only the large path difference points in the interferogram signals (ignoring the few points at zero path difference where the photon statistics depends upon the relative phase between the two combs), the quantum-limited noise in the two discretely-sampled interferograms (reference and gas) can thus be represented as discrete gaussian random processes. The processes are white (each time sample is independent) but there is a fixed amount of correlation between simultaneous samples from both processes. These two correlated discrete processes can be generated from two independent sequences and a random variable transformation to induce any desired correlation $\rho$.

Therefore, $x[n]$ and $y[n]$ are defined as two zero mean independent discrete random processes such that their auto and cross-correlations are $R_{xx}[n] = R_{yy}[n] = \sigma^2\delta[n]$ and $R_{xy}[n] = 0$, where $n$ is the interferogram point index number. The sequences $u[n]$ and $v[n]$ corresponding to the noise on each measured photocurrent are obtained from the transformation:

$$u[n] = x[n]$$

$$v[n] = \rho x[n] + \sqrt{1-\rho^2}\, y[n]$$

The two new sequences $u[n]$ and $v[n]$ are zero mean, have the same variance $\sigma^2$ and autocorrelation as the original sequences $x[n]$ and $y[n]$ but now have the specified correlation $\rho$ for the same sample index $n$. The correlation coefficient $\rho$ is influenced by the amount of squeezing but can also be adjusted to account for independent additive noise sources in each detection channel. The spectrum is estimated by performing a discrete Fourier transform (DFT) on a measured discrete interferogram. For instance, we can obtain the complex discrete spectrum $X[k]$ of the sequence $x[n]$ from:

$$X[k] = X_r[k] + iX_i[k] = \sum_{n=0}^{N-1} x[n] e^{-i\frac{2\pi}{N}kn},$$

where $k$ is the spectral bin index, $N$ is the total number of measured points and $X_r[k]$, $X_i[k]$ are respectively the real and imaginary parts of the spectrum. Since the DFT is a linear operator, it is straightforward to show that $X[k]$ and $Y[k]$ are zero-mean, independent and have a variance equal to $N\sigma^2$. The real and imaginary parts of each discrete spectral noise are also zero mean, independent and have variance $N\sigma^2/2$. Again, by virtue of the Fourier transform linearity, one writes:

$$U[k] = X[k],$$

$$V[k] = \rho X[k] + \sqrt{1-\rho^2}Y[k].$$

Thus, $U[k]$ and $V[k]$ are zero mean, have a variance $N\sigma^2$ and have a cross-correlation $R_{UV}[k] = \rho\,\delta[k]$.

Spectral Transmittance of Correlated Gaussian Random Processes

The desired measurement in spectroscopy is often the spectral transmittance of weak absorbing species $T[k]$ which spectrally filters the source spectrum $S[k]$. A complex ratio between measurements with and without the absorber aims to remove the contribution from $S[k]$ but both measurements are altered by noise. After the beam splitter, the two signals have opposite signs so that the two measurements are $M_1[k] = S[k]T[k] + V[k]$ and $M_2[k] = -S[k] + U[k]$. The measurement model is thus:

$$M[k] = \frac{M_1[k]}{-M_2[k]} = \frac{S[k]T[k] + V[k]}{S[k] - U[k]}.$$

(1)

We assume here that filtering the optical spectrum $S(\nu)$ by the transmittance $T(\nu)$ does not change the detected photocurrent statistics. This assumption is not true in a fully quantum mechanical description of the measurement. This model is valid for weak absorption where the squeezing is largely preserved despite the optical loss and in line with the hypothesis that for most interferogram delays, the measurement is intensity-like and free induction signal is small. Performing a series expansion on Equation 1 gives:

$$M[k] \approx T[k] + \frac{V[k]}{S[k]} + T[k]\frac{U[k]}{S[k]}.$$

(2)

And subsequently:

$$M[k] \approx T[k] + (\rho + T[k])\frac{X[k]}{S[k]} + \frac{\sqrt{1-\rho^2}Y[k]}{S[k]} + \mathcal{O}\left(\left(\frac{X[k]}{S[k]}\right)^2\right).$$

(3)

Equation 3 shows that two independent noise sources $X[k]$ and $Y[k]$ are scaled by $1/S[k]$, as is routinely seen in transmittance measurements: the signal to noise ratio degrades where signal is low. This effect is visible in Figure 6 in the main paper. It is also seen that as the correlation $\rho \to -1$ ($U[k]$ and $V[k]$ are anti-correlated), the noise term proportional to $Y[k]$ decreases while the magnitude of the term proportional to $X[k]$ also decreases if $T[k] \approx 1$ (weak absorption) but increases if $T[k] \to 0$. This analysis shows that in the limit of small absorption, the complex ratio of two correlated gaussian random processes yields a signal that retains the correlation of the individual processes.

Shot noise in squeezed dual-comb spectroscopy

Here, we heuristically model quantum noise reduction in dual-comb spectroscopy. As seen from the previous section, the noise of a temporal mode-mismatched measurement resembles the noise generated by the addition of photocurrents detected at the two beamsplitter ports in a DCS experiment. Thus, it is sufficient to consider DCS SNR for a single photodetector. We define $P_1$ and $P_2$ as the optical powers of comb 1 and comb 2 on a single photodetector. We assume there is no dynamic range limitation and that shot noise is the dominating noise contribution which is a safe assumption for optical powers in the milliwatt range with relative intensity noise at the -160 dBc/Hz. In this case, the power dependence of the shot-noise-limited SNR, $R_{SNL}$ in a measurement-specified bandwidth, can be written as:

$$R_{SNL} \propto \frac{\sqrt{P_1 P_2}}{\sqrt{P_1 + P_2}}.$$

For the power-balanced ($P_1 = P_2 = P_{TOT}/2$) case this expression reduces to:

$$R_{SNL} \propto \sqrt{P_{TOT}} \propto \sqrt{N_{TOT}},$$

where $N_{TOT}$ is total number of detected photons. If $P_1 \gg P_2$ we can write:

$$R_{SNL} \propto \sqrt{P_2}.$$

Thus, in the power-unbalanced case for coherent states, the SNR depends solely on the power of weak comb. In the demonstrated scenario of amplitude-squeezed dual-comb spectroscopy, amplitude squeezing reduces the noise on the stronger comb below the SNL. In this case we have:

$$R_{SQZ} = (1/\gamma_1)\, R_{SNL},$$

where $0 < \gamma_1 \leq 1$ is the noise reduction factor due to squeezing. In our experiment, we demonstrated $\gamma_1 \approx 1/\sqrt{2}$ as the noise reduction factor on the shot noise standard deviation. For the case where both combs are amplitude-squeezed, we can write:

$$R_{SQZ}^2 \propto \frac{P_1 P_2}{\gamma_1^2 P_1 + \gamma_2^2 P_2}.$$

In the power-balanced case, the SNR is maximized for a fixed total power. In this case, the SNR can be rewritten in terms of the coherent state SNL as follows:

$$R_{SQZ}^2 = R_{SNL}^2 \left(\frac{2}{\gamma_1^2 + \gamma_2^2}\right).$$

For $\gamma_1 = 1/\sqrt{2}$ and $\gamma_2 = 1$ this expression gives $R_{SQZ}^2 = 1.33\, R_{SNL}^2$. Even with infinite squeezing on one comb (i.e. $\gamma_1 \to 0, \gamma_2 = 1$), the advantage in SNR is maximally $R_{SQZ}^2 = 2\, R_{SNL}^2$. For $\gamma_1 = \gamma_2$ this equation reduces to $R_{SQZ} = (1/\gamma_1)\, R_{SNL}$ as in the un-balanced case.

Figs. S1 to S5

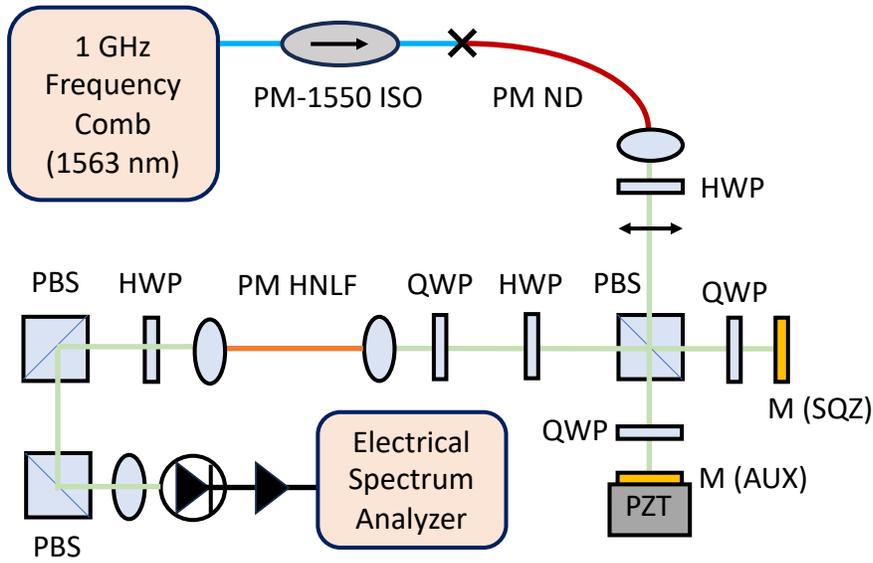

**Fig. S1. Schematic of single-comb squeezing experiment.** A single 1 GHz frequency comb is compressed and injected into a birefringence compensator stage which generated two orthogonally polarized pulses. These pulses are coupled into a PM-HNLF and then recombined in free space after nonlinear propagation. The combined pulses are detected by a single photodiode. The electrical output is then amplified and characterized by an electrical spectrum analyzer.

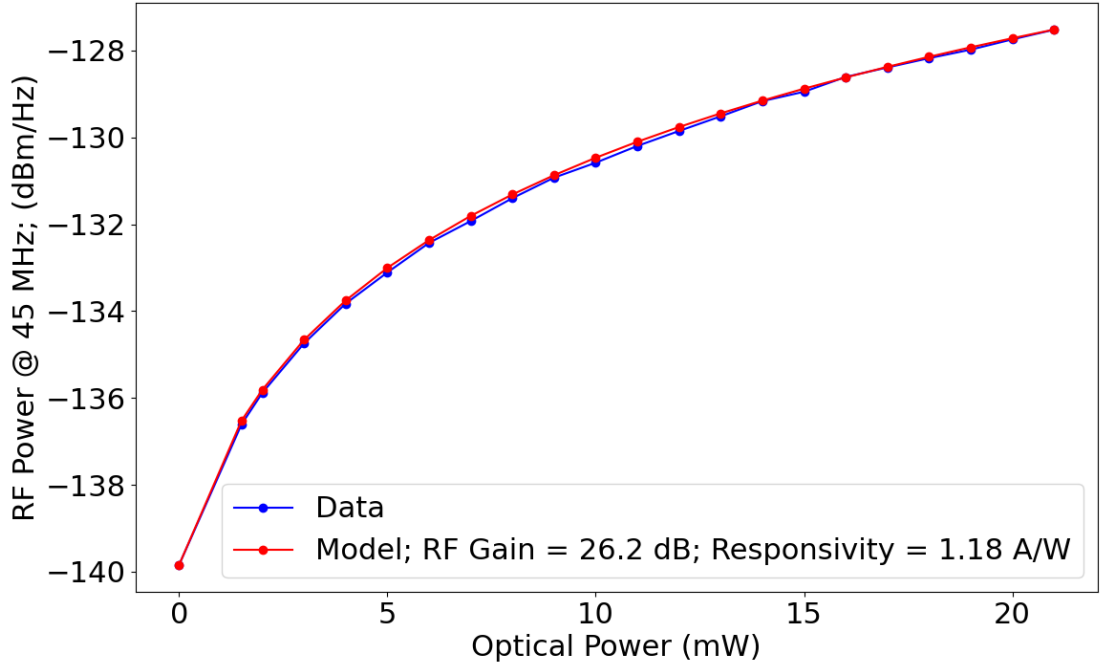

**Fig. S2. Shot noise calibration for the high quantum efficiency detector.** A known value of optical power is sent to the Fermionics (high quantum efficiency) diode. The noise is recorded on an electrical spectrum analyzer with span 10 MHz to 100 MHz and resolution/video bandwidth both set to 1 kHz. A narrow band at 45 MHz is integrated to get one noise value as the power is swept. The manufacturer responsivity is used (1.18 A/W) and the radio frequency gain of the detection circuit is independently estimated to be 26.2 dB. The measured detection circuit noise floor is -139.8 dBm/Hz. The measured dataset is a very close match to the shot noise with noise floor model which both indicates that our combs are shot noise limited in the vicinity of 45 MHz and our detector is not in saturation.

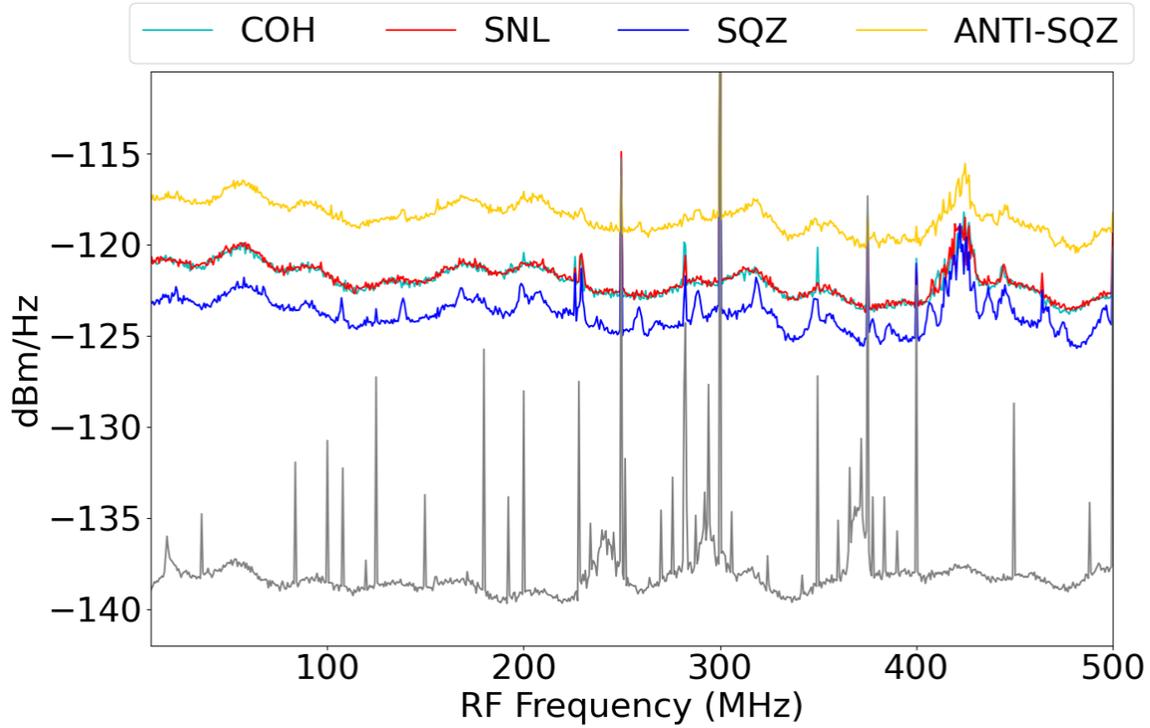

**Fig. S3. Squeezing with a low quantum efficiency, high-speed photodetector.** Broadband squeezing of the 1 GHz comb is measured using a Thorlabs BDX1BA photodiode (single port of balanced detector). Several peaks are attributed to the GAWBS effect (*e.g.* 23 MHz, 107 MHz, 139 MHz, 168 MHz,199 MHz etc.). Noise at 420 MHz seems to be amplitude noise on the comb itself. Other narrow spikes are RF interference. Sinusoidal modulation of the RF noise is due to lack of 50 Ohm transimpedance element directly after the photodiode in this specific measurement. Electrical noise floor is shown in gray.

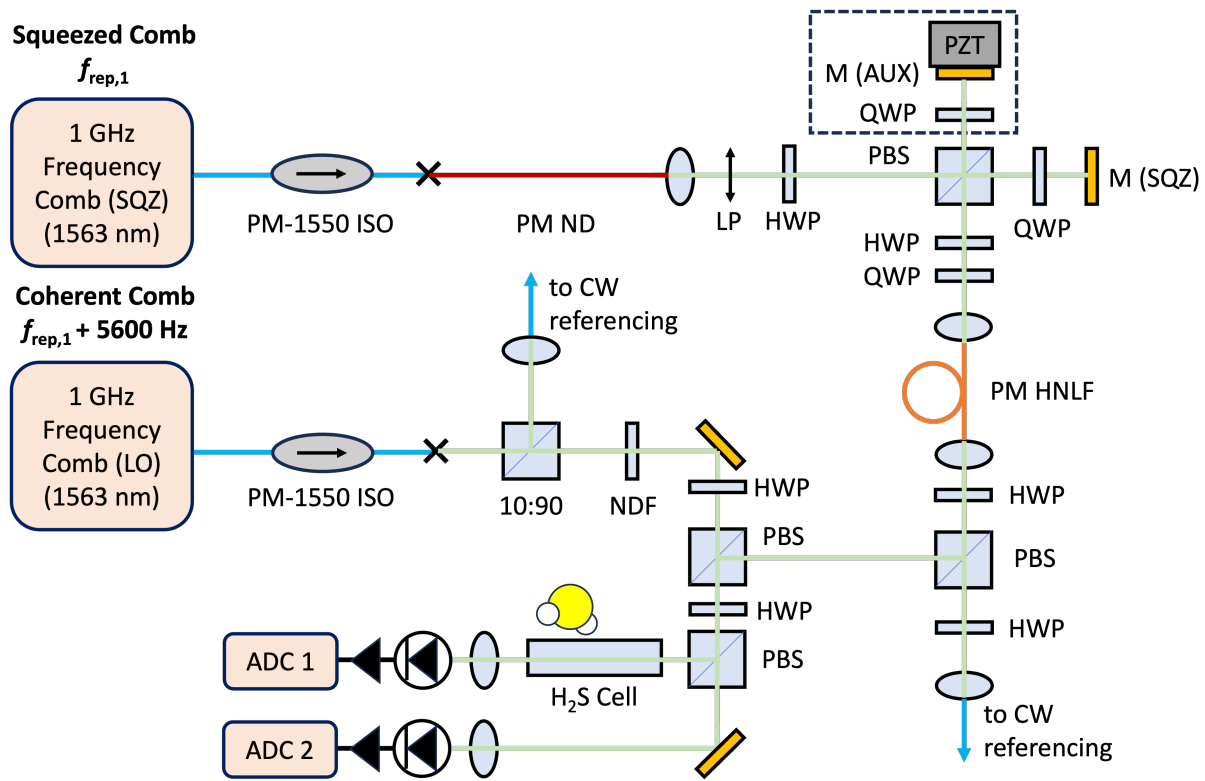

**Fig. S4. Detailed schematic of quantum-enhanced dual-comb experiment.** Full schematic including squeezed comb generation, optical reference pick-off points and dual-comb absorption spectroscopy setup. PM-1550 ISO = polarization-maintaining 1550 nm fiber isolator; PM ND = polarization-maintaining normal dispersion fiber; LP = linear polarizer; HWP = half wave plate; QWP = quarter wave plate; M = mirror; PZT = piezo-electric transducer; PM HNLF = polarization-maintaining highly nonlinear fiber; PBS = polarizing beam splitter; 10:90 = 10% (90%) reflection (transmission) beam splitter; NDF = neutral density filter; ADC = analog-to-digital converter. Small oval indicates fast aspheric lens.

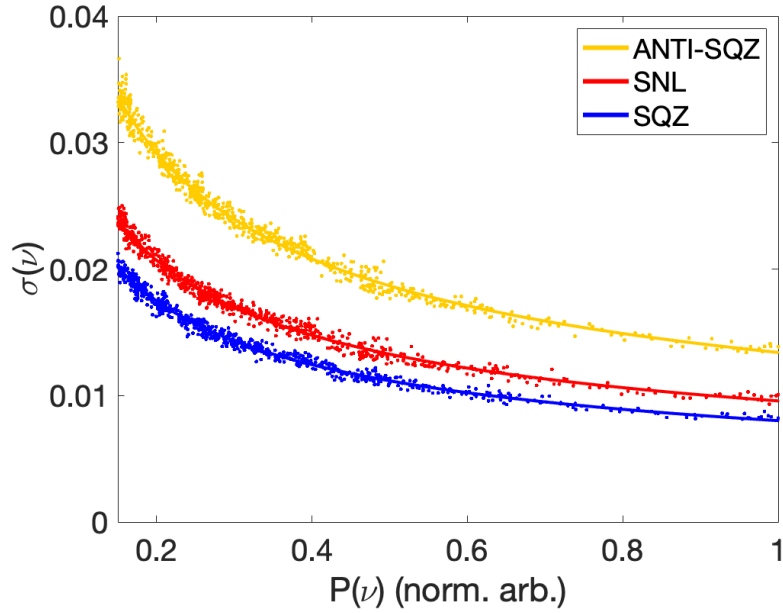

**Fig. S5. Shot noise of DCS residuals.** By fitting each individual transmission spectrum to the HITRAN model, we extract time-resolved spectral residuals over a series of 1100 sample spectra. The standard deviation of the residuals for each comb mode, $\sigma(\nu)$, is calculated along the time axis. This value is plotted against the mean power for the comb mode $P(\nu)$. As expected, all three measurement cases show a $\sqrt{P}$ dependance but with different pre-factors (*16*).

31